\documentstyle[twoside,fleqn,npb,epsfig]{article}
\newcommand{\AmS}{{\protect\the\textfont2
  A\kern-.1667em\lower.5ex\hbox{M}\kern-.125emS}}
\newcommand{\etal}{{\it et al.}}

\newcommand{\pl}{Phys. Lett. }

\title{Polarized lepton nucleon scattering ---\\
  summary of the experimental spin sessions at DIS~99}

\author{Michael D\"uren\\
   Physikalisches Institut, Universit\"at Erlangen-N\"urnberg,
   91058 Erlangen, Germany\\
 {\small E-mail: {\tt Michael.Dueren@desy.de} }}

\begin{document}

\begin{abstract}

  This paper summarizes the contributions to the experimental sessions
  on polarized lepton nucleon scattering at the DIS 99 workshop.
  Results are reported about the flavor decomposition of the quark
  polarization, a first direct measurement of a positive gluon
  polarization, the observation of a double-spin asymmetry in
  diffractive $\rho^0$ production, the polarization of $\Lambda$
  hyperons, the observation of transverse single-spin asymmetries and
  the measurement of the Gerasimov-Drell-Hearn sum rule. Prospects of
  future fixed target and collider facilities are discussed.

\end{abstract}
\maketitle

\section{OVERVIEW}

Exciting and even unexpected results have been presented in the
sessions about ``polarized lepton nucleon scattering'', at the DIS~99
workshop. Most striking are the presentations of the {\em flavor
  decomposition of the quark polarization}, the first direct
observation of a {\em positive gluon polarization} in the nucleon and
the observation of a {\em double-spin asymmetry in diffractive
  $\rho^0$ production}. First results on the Gerasimov-Drell-Hearn
(GDH) sum rule have been reported.  The interesting and new fields of
polarization phenomena related to transversity, leading twist-3
distributions and fragmentation came into the reach of
experimentalists this year.  Promising results were announced about
azimuthal pion asymmetries and the polarization of $\Lambda$ hyperons
in the final state.

\section{INCLUSIVE SPIN PHYSICS}

Spin structure functions have been measured for many years, and an
impressive, precise data set has been collected by the experiments at
SLAC, by SMC and by HERMES~\cite{Emlyn}. At all three sites the spin
structure function $g_1(x,Q^2)$ was measured for the proton and the
neutron and consistent results were obtained that agree with the $Q^2$
evolution as predicted by QCD. The experiment E155x at SLAC is running
in spring 1999 with the aim to provide precise data on $g_2$ and to
gain access to the interesting twist-3 component of $g_2$ that remains
after subtraction of the Wandzura-Wilczek contribution $g_2^{WW}$.

The situation of the spin sum rules has not changed significantly
since the last DIS meeting. The Ellis-Jaffe sum rule is found to be
violated and all results support the Bjorken sum rule. The precision
of the sum rule tests is limited by theoretical uncertainties in the
extrapolation of the spin structure functions at low x.
Experimentally, the low x range will be accessible only by a future
high energy spin experiment at a new facility as the proposed
polarized HERA collider~\cite{Hughes}.  A measurement of the high $x$
region at low energies is planned at Jefferson Lab~\cite{Meziani}.

\subsection{Gerasimov-Drell-Hearn sum rule}

\begin{figure}[htb]
\epsfxsize 70mm {\epsfbox{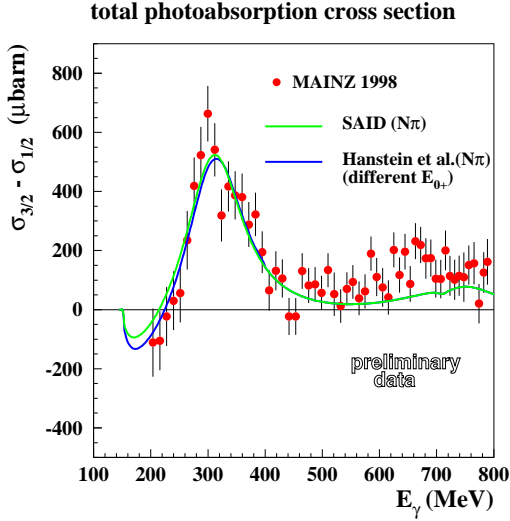}}
\caption{The spin dependent part of the total photoproduction cross section
  as a function of the photon energy as measured at Mainz.}
\label{fig:gdhmainz}
\end{figure}

The GDH sum rule relates the polarization dependent part of the total
photoproduction cross section to the anomalous magnetic moment of the
nucleon. This important relation has been tested by a precision
experiment at the tagged polarized photon beam of the microtron MAMI
in Mainz, Germany. The energy range accessible by the experiment is
200-800~MeV.  Fig.~\ref{fig:gdhmainz} shows a preliminary analysis of
a small subset of the data.  For the integral a number of $230\pm 20
~\mu$b was reported, which accounts for most of the GDH prediction of
204 $\mu$b~\cite{Thomas}. The remaining difference might be due to
contributions at higher energies which are planned to be measured at
Bonn.

The GDH sum rule can be generalized for electroproduction. Data have
been recently published by the HERMES collaboration~\cite{gdh} and a
new measurement in the important resonance region was performed
in spring 1999 as experiment E94-010 at Jefferson Lab~\cite{Meziani}.
First raw data of this experiment have been shown at this conference.

\section{SEMI-INCLUSIVE DIS}
\subsection{Flavor decomposition}

The violation of the Ellis-Jaffe sum rule, which was first reported by
EMC~\cite{emc1}, was interpreted as evidence for the fact that only a
fraction of the nucleon spin can be attributed to the quark spins and
that the strange quark sea seems to be negatively
polarized~\cite{elkarl}. This caused the so-called {\em spin crisis}.
This interpretation is based on the assumption that the spin structure
of the baryon multiplets is SU(3)$_f$ flavor symmetric.

Semi-inclusive data can be used to measure the sea polarization
directly and to test SU(3)$_f$ symmetry by comparing the first moments
of the flavor distributions to the SU(3)$_f$ predictions. In addition,
semi-inclusive polarized DIS experiments can determine the
separate spin contributions $\Delta q_f$ of quark and antiquark
flavors $f$ to the total spin of the nucleon not only as a total
integral but as a function of the Bjorken scaling variable $x$.

Hadron production in DIS is described by the absorption of a virtual
photon by a point-like quark and the fragmentation into a hadronic
final state. The two processes can be characterized by two functions:
the quark distribution function $q_f(x,Q^2)$, and the fragmentation
function $D_f^h(z,Q^2)$.  The semi-inclusive DIS cross section
$\sigma^h(x,Q^2,z)$ to produce a hadron of type $h$ with energy
fraction $z=E_h/\nu$ is then given by
\begin{equation}
    \label{eq:hadronsig}
    \sigma^h(x,Q^2,z)\propto\sum_f e_f^2 q_f(x,Q^2) D_f^h(z,Q^2).
\end{equation}
 In the target rest frame, $E_h$ is the energy of the
hadron, $\nu=E-E'$ and $-Q^2$ are the energy and the squared
four-momentum of the exchanged virtual photon, $E$($E'$) is the energy
of the incoming (scattered) lepton and $e_f$ is the quark charge in
units of the elementary charge.  The Bjorken variable $x$ is
calculated from the kinematics of the scattered lepton according to
$x=Q^2/2M\nu$ with $M$ being the nucleon mass.  It is assumed that the
fragmentation process is spin independent, i.e.~that the probability
to produce a hadron of type $h$ from a quark of flavor $f$ is
independent of the relative spin orientations of quark and nucleon.
The spin asymmetry $A_1^h$ in the semi-inclusive cross section for
production of a hadron of type $h$ by a polarized virtual photon is
given by
\begin{equation}
      A_1^h(x,Q^2,z)= C_R{\sum_f e_f^2  \Delta q_f(x,Q^2) D_f^h(z,Q^2)\over
      \sum_f e_f^2 q_f(x,Q^2)D_f^h(z,Q^2)}
    \label{eq:hadronasym}
\end{equation}
where $\Delta q_f(x,Q^2)=
q_f^{\uparrow\uparrow}(x,Q^2)-q_f^{\uparrow\downarrow} (x,Q^2)$ is the
polarized quark distribution function and
$q_f^{\uparrow\uparrow(\uparrow\downarrow)}(x,Q^ 2)$ is the
distribution function of quarks with spin orientation parallel
(anti-parallel) to the spin of the nucleon.  The unpolarized quark
distribution functions are defined by $F_2$ (and not by $F_1$):
\begin{equation}
  \label{eq:f2}
  F_2=\sum_f e_f^2  x q_f(x,Q^2)
\end{equation}
and they include therefore a longitudinal component of the photon
absorption cross section.
 The term 
\begin{equation}
    C_R={1+R(x,Q^2)\over 1+\gamma^2},
    \label{eq:hadronc}
\end{equation}
with $R=\sigma_L/\sigma_T$ being the ratio of the longitudinal to
transverse photon absorption cross section, corrects for the
longitudinal component which {\it a priori} is not present in the
asymmetry and the polarized distribution functions.  It is assumed
that the ratio of longitudinal to transverse components is flavor and
target independent and that the contribution from the second spin
structure function $g_2(x,Q^2)$ can be neglected.  The term
$\gamma=\sqrt{Q^2}/\nu$ is a kinematic factor which enters from the
$g_2=0$ assumption.  Eq.~(\ref{eq:hadronasym}) can be used to extract
the quark polarizations $\Delta q_f(x)/q_f(x)$ from a set of measured
asymmetries on the proton and neutron for positively and negatively
charged hadrons.

\begin{figure}[htb]
\epsfxsize 80mm {\epsfbox{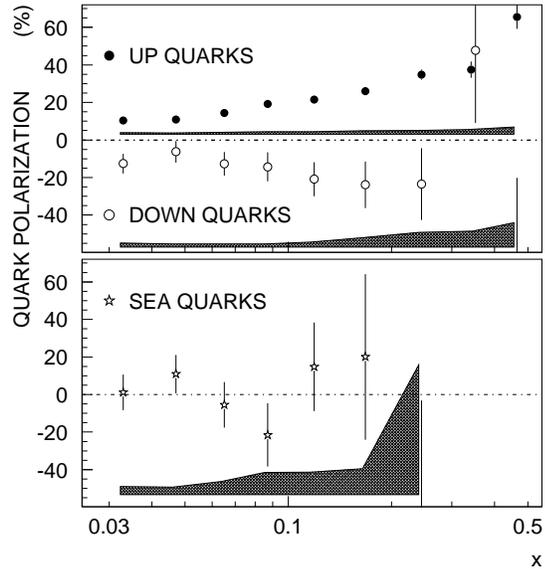}}
\caption{The polarization of quarks in the proton has been measured by
  HERMES as a function of $x$, separately for the flavors up and down and
  for the sea. The error bars shown are the statistical and the
  bands the systematic uncertainties.}
\label{fig:dis99pol}
\end{figure}

Results on the decomposition of the proton spin into contributions
from the valence spin distributions $\Delta u_v$ and $\Delta d_v$ and
from the sea $\Delta q_s$ have been previously reported by
SMC~\cite{smcsemi}. New, more precise data from HERMES have been
presented at this workshop~\cite{Ruh}. Fig.~\ref{fig:dis99pol} shows
the polarization $\Delta q/q$ of quarks in the proton, separated into
flavors. The {\em up} flavor has a positive polarization which reaches
about 40\% at large $x$ whereas the {\em down} flavor has a
polarization opposite to the proton spin, in excess of 20\%. In the
sea region at small $x$ the up and down polarizations do not vanish
completely. The sea polarization itself is compatible with zero as
shown in the lower panel. The extraction of the sea was done under the
assumption that the polarization of the sea quarks is independent of
their flavor.

%

\begin{figure}[htb]
\vspace{9pt}
\epsfxsize 75mm {\epsfbox{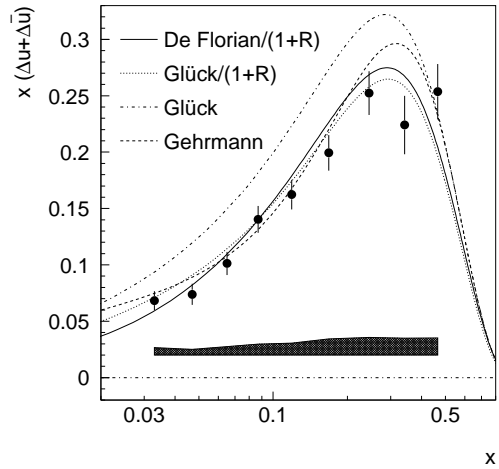}}
\caption{
  The quark spin distribution $x(\Delta u(x)+\Delta \bar{u}(x))$ for
  $Q^2=2.5$~GeV$^2$ as measured by HERMES is compared to different
  sets of parametrizations which correspond to the following
  publications: De Florian {\it et al.}~($0.1<\Delta G<0.8$,
  LO)~\protect\cite{Flori}, {Gehrmann and Stirling} ('Gluon A',
  LO)~\protect\cite{GER96}, and Gl\"uck {\it et al.}~('Standard
  Scenario', LO)~\protect\cite{GLU96}.  The {De Florian} and {Gl\"uck}
  parametrizations are corrected by a factor $(1+R)$ to allow for a
  direct comparison. The error bars shown are the statistical and the
  band the systematic uncertainty.}
  \label{fig:dis99dis}
\end{figure}
The polarized quark distribution functions $\Delta q_f(x)$ have been
extracted by multiplying the quark polarizations $\Delta
q_f(x)/q_f(x)$ with the known unpolarized distribution functions.
Fig.~\ref{fig:dis99dis} shows that the HERMES result agrees with the
LO parametrization of Ref.~\cite{GER96} of world data. However, not
all LO parametrizations agree with HERMES and it seems that the
parametrizations are internally inconsistent.  Agreement can be
achieved by dividing out a factor (1+R).  The explanation is that the
ratio R of the longitudinal to transverse cross section is usually set
to zero in the LO fits. On the other hand experimentalists usually use
the measured R, which is non-zero, to extract $g_1$. The result of the
fit then depends on the choice of the input for the fit, either $A_1$
(as in Ref.~\cite{Flori,GLU96}) or $g_1$ (as in Ref.~\cite{GER96}).
From the relation
\begin{equation}
    \label{eq:a1}
  A_1 = {2xg_1\over F_2}{(1+R)\over (1+\gamma^2)},
\end{equation}
which is quoted here for the approximation that $g_2=0$, it follows
directly that the two choices will give different results.

\begin{table}[ht]
  \caption{
    The integrals of various spin distributions as measured by HERMES
    for  $Q^2=2.5$~GeV$^2$. 
    Note that the entry for $\Delta s +\Delta \bar{s}$ does not represent a
    direct measurement of the strange sea but relies on the assumption
    that the sea polarization is flavor independent.
    An uncertainty of the Regge-type extrapolation at low $x$
     is not included in the quoted error.}
   \label{tab:moment}
\begin{center}
  \begin{tabular}{|c||r|}
    \hline
    & \multicolumn{1}{c|}{total integral}\\
    \hline\hline
    $\Delta u +\Delta \bar{u}$ &
    $ 0.56\pm 0.02\pm 0.03 $\\
    $\Delta d +\Delta \bar{d}$ &
    $-0.25\pm 0.06 \pm 0.05$ \\ 
    $\Delta s +\Delta \bar{s}$ &
    $-0.02\pm     0.03\pm 0.04 $ \\
    \hline
    $\Delta q_0$ &
    $  0.28\pm 0.04\pm 0.09                     $   \\
    $\Delta q_3$ &
    $ 0.83\pm     0.07\pm 0.06                           $  \\
    $\Delta q_8$ &
    $  0.32\pm     0.09\pm 0.10                          $  \\
    \hline
    $\Delta u_{v}$ &
    $ 0.57\pm 0.05\pm 0.08                      $ \\
    $\Delta d_{v}$ &
    $-0.21\pm 0.10\pm 0.13                      $\\
    \hline
    $ x\Delta u_{v}$ &
    $ 0.12 \pm 0.01 \pm 0.01  $\\
     $ x \Delta d_{v}$ &
    $-0.02 \pm 0.02 \pm 0.02  $ \\
    \hline
  \end{tabular}
\end{center}
\end{table}
The first and second moments of the spin distributions have been
determined by HERMES. In the measured region, the results agree
between HERMES and SMC within the quoted errors.  A simple Regge-type
extrapolation has been applied at low $x$ to obtain the total integrals
as quoted in Table~\ref{tab:moment}. The HERMES results for the first
and second moment of $\Delta u_v$ show a significant discrepancy with
a prediction from quenched lattice QCD in Ref.~\cite{goeck}. The
result for $\Delta u +\Delta \bar u$ is inconsistent with the result
from the inclusive data based on SU(3)$_f$ flavor symmetry as in
Ref.~\cite{elkarl}. The inconsistency of the up flavor has its
counterpart in the difference which is observed in the sea results.
The inclusive analysis obtains a large negative strange sea compared
to the zero sea in the semi-inclusive analysis. Possible explanations
of these differences are that either SU(3)$_f$ is violated, which
would modify the inclusive result, or that the assumption about the
flavor independence of the sea is wrong, which would modify the
semi-inclusive result.  To test the applicability of SU(3)$_f$ and
SU(2)$_f$ flavor and isospin symmetry, the semi-inclusive results for
$\Delta q_8$ and $\Delta q_3$ have been compared to the predictions
$\Delta q_8=3F-D$ and $\Delta q_3=F+D$ (Bjorken sum rule). Both
predictions agree with the HERMES results when the appropriate QCD
corrections are taken into account.  For a decisive conclusion about
the origin of the discrepancy, the precision has to be further
improved and the sea assumption has to be tested explicitly.

A significant improvement of the precision of the $\Delta d(x)+\Delta
\bar d$ determination is expected in near future from HERMES using the
1999 deuterium data set.  The newly installed RICH will allow a direct
measurement of $\Delta s(x)$ using Kaon identification.

\subsection{The gluon polarization}

The ``most wanted'' component of the nucleon spin is the polarization
of gluons, as they are probably responsible for the spin deficit of the
quarks. As the virtual photon does not couple directly to gluons, a
measurement of the gluon polarization was up to now only very
indirectly possible by using the QCD evolution equations which relate
the $Q^2$-dependence of the quark distributions to the gluon
distribution.

For the first time a more direct measurement of the gluon polarization
has been announced~\cite{Amarian}. By selecting events with two
hadrons with opposite charge and with large transverse momentum,
HERMES was able to accumulate a sample of events which is enriched by
photon-gluon fusion events. By requiring a large transverse momentum
of 1.5~(1)~GeV/c for the first (second) hadron, the sub-process where
the gluon splits into two quarks has a hard scale and can be treated
pertubatively.  HERMES estimates from Monte-Carlo studies that the
average squared transverse momentum of the quarks is 2.1~(GeV/c)$^2$.
As long as the fragmentation process is spin independent, the spin
asymmetry in the production of the quark-antiquark pair is the same as
the spin asymmetry of the observed final state. The measured asymmetry
is however affected by background processes. The unique signature of
the HERMES result is that the spin asymmetry comes out negative.  All
background processes have a positive asymmetry, as long as they are
dominated by the positive polarization of the ``up''-quarks in the
proton. The observed negative asymmetry can be explained by a
significant positive gluon polarization. The change of sign comes from
the negative analyzing power of the photon-gluon fusion diagram.
Using a specific background Monte Carlo, HERMES obtains a value of the
gluon polarization of $\Delta G/G=0.41\pm 0.18\pm 0.03$ at $x_G=0.17$.
The quantitative result depends however critically on the detailed
understanding of the background processes.

\subsection{Transverse asymmetries}

The next step in polarized DIS beyond the understanding of the
collinear part of the quark and gluon polarization in the nucleon is
the understanding of the transverse polarization components.
Single-spin asymmetries in polarized hadronic reactions are
interpreted as effects of {\em time-reversal-odd} distribution
functions (Sivers mechanism) or {\em time-reversal-odd} fragmentation
functions (Collins mechanism).  The numerous possible processes were
discussed in the theory spin sessions~\cite{Vogelsang}.

SMC presented at this conference the first measurement of
semi-inclusive DIS hadron production on a transversely polarized
target~\cite{Bravar}. Leading hadron production has been analyzed in
terms of the Collins angle and indeed a non-zero asymmetry
$A_N=11\%\pm 6\%$ has been found for positive hadrons, whereas the
negative hadrons yield $-2\%\pm 6\%$.

\begin{figure}[htb]
\vspace{9pt}
\epsfxsize 80mm {\epsfbox{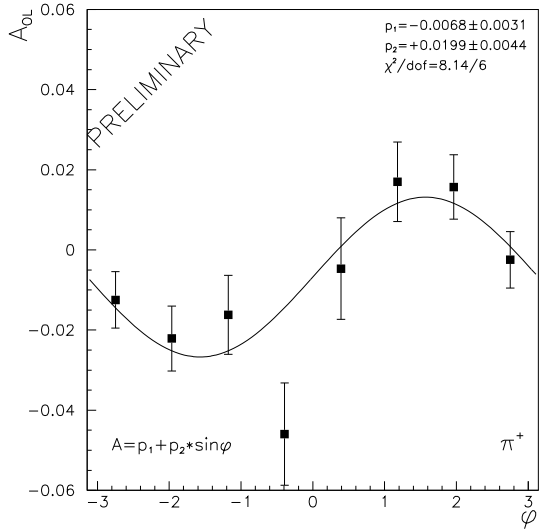}}
\caption{Azimuthal dependence of the single-spin asymmetry in the cross section
  for $\pi^+$. The error bars are statistical uncertainties.}
    \label{fig:azim}
\end{figure}

A significant result has been reported by HERMES on a related
quantity~\cite{Avakian}. HERMES measured the asymmetry of hadron
production on a longitudinally polarized target. Even in this case an
asymmetry is expected in the azimuthal angle between the plane which
contains the produced pion and the virtual photon and the plane which
contains the scattered lepton and the virtual photon.
Fig.~\ref{fig:azim} shows this single-spin asymmetry as a function of
the azimuthal angle for positive pions. A sinusoidal fit yields an
asymmetry of $A_N=2.0\%\pm 0.4\% $ for the positive and
$A_N=-0.1\%\pm 0.5\% $ for the negative pions. The good
statistical precision of this result is due to the hadron acceptance
of the detector and the pure, highly polarized hydrogen target.

%
\subsection{$\Lambda$ polarization}

A related quantity is the polarization of $\Lambda$ hyperons. HERMES
reported two results here~\cite{Belostotski}.

The first one is the measurement of the polarization transfer in DIS
scattering of longitudinally polarized electrons off unpolarized
targets. A $\Lambda$ polarization of $P_\Lambda=0.03\pm 0.06 \pm 0.03$
was reported, a number which is not precise enough to distinguish
between different predictions. A naive quark model which assumes 100\%
polarization of s-quarks in $\Lambda$ hyperons predicts
$P_\Lambda=0.018$, whereas a SU(3)$_f$ symmetric model from Jaffe
predicts $P_\Lambda=-0.057$.

\begin{figure}[htb]
\vspace{9pt}
\epsfxsize 80mm {\epsfbox{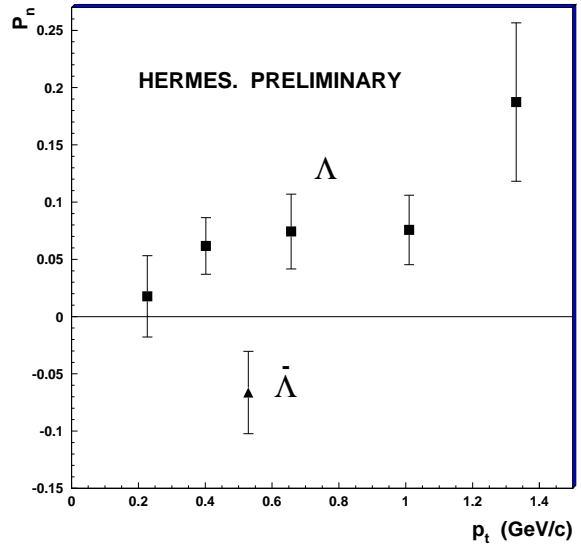}}
\caption{The squares (triangles) indicate the  transverse
  polarization of $\Lambda$ ($\bar \Lambda$) in unpolarized
  photoproduction as function of the transverse momentum. }
    \label{fig:lambda}
\end{figure}
A much more precise result was reported concerning the transverse
polarization of $\Lambda$ hyperons in quasi-photoproduction off an
unpolarized target:
\begin{equation}
  \label{eq:lambda}
  \gamma^{(*)}p\to \Lambda X.
\end{equation}
The polarization was measured in respect to the plane perpendicular to
the $\Lambda$ production plane.  Fig.~\ref{fig:lambda} shows the
polarization as a function of the transverse momentum of the $\Lambda$
and $\bar \Lambda$.  A large positive polarization is observed for
$\Lambda$ hyperons, with the tendency to increase with its transverse
momentum. The $\bar \Lambda$ antihyperons show a negative
polarization.  There is no straight-forward explanation of the
observed asymmetries in QCD; however, similar polarizations have been
found in hadronic collisions.

\section{DIFFRACTIVE DOUBLE-SPIN ASYMMETRIES}

Results on double-spin asymmetries in diffractive $\rho^0$-production
have been reported by SMC~\cite{Tripet} and HERMES~\cite{Meissner}.
Naively, no spin asymmetry is expected in the approach where
diffraction is described by the exchange of a pomeron with vacuum
quantum numbers. In this frame there should be no way that the
diffractive $\rho^0$ production knows about the spin of the target
nucleon. The SMC result is in agreement with this prediction. No
significant asymmetry has been observed.

HERMES reported a significant, unexpected positive asymmetry of
$A_1^\rho=0.30\pm 0.11 \pm 0.05$.  The asymmetry in the production of
other vector mesons $\phi$ and $J/\psi$ was also measured, but with
much less precision and came out to be compatible with zero. The
non-zero $\rho$-asymmetry can possibly be understood by comparing this
process $\gamma^*+p\to p+\rho$ to the similar process $\gamma^*+p\to
p+\gamma^*$ which is related to deep inelastic scattering and its well
known positive asymmetry $A_1^p$. The difference between the HERMES
and SMC result may have its reason in the different $Q^2$ and $W^2$
ranges of the two experiments.

\section{FUTURE FACILITIES}

The future of polarized DIS will consist of both, fixed target~\cite{Kabuss}
and  collider experiments~\cite{Hughes,GotoSaito}.

\subsection{Fixed target}

At SLAC the fixed target inclusive era will end with the precise
measurement of $g_2^{p,d}$ at E155x. At lower energies, MAMI at Mainz,
ELSA at Bonn and CEBAF at Jefferson Lab will continue to do spin
physics. The main future players at higher energies will
be HERMES at DESY and COMPASS at CERN. Both experiments will
concentrate on semi-inclusive results.

The main aims of the experiments are the measurement of the gluon
polarization, the flavor decomposition of polarized quark
distributions, polarized vector meson production, polarized
fragmentation functions, transversity, and in the case of COMPASS also
the angular momentum of quarks and gluons and off-forward parton
distributions.

HERMES has an upgrade program to improve particle identification with
pion, kaon and proton separation in the full kinematic region, using a
RICH detector. An improved muon acceptance and identification will
allow for a better $J/\Psi$ detection. A wheel of silicon detectors
just behind the target cell improves the acceptance
especially for $\Lambda$ decay products. A recoil detector system is
dedicated to low energy, large angle target fragments and spectator
nucleons.

COMPASS will start in 2000 in the experimental area where the SMC
experiment has been, however with an improved beam, improved target,
high luminosity and, compared to the SMC experiment, with a much
better and larger hadron acceptance and particle identification. In
the final stage COMPASS will have two spectrometer magnets, two RICH
detectors, two hadron and two electromagnetic calorimeters.  Compared
to HERMES, the large beam energy of COMPASS of 100-200~GeV enables
measurements at smaller $x$ and larger $Q^2$ and $W^2$. The large
$W^2$ allows for charm production well above the threshold and for the
generation of hadrons with large transverse momentum. The production
of open charm allows for a direct measurement of the gluon
polarization.

A future fixed target machine is ELFE, a possible new European
electron machine in the 15-30 GeV range, which is discussed in
connection with the TESLA project at DESY and also at CERN as a
machine which could re-use the cavities from LEP. Two further
experiments which both proposed to measure the gluon polarization via
charm production in photoproduction were proposed some years ago,
 and are still discussed: E-156 at SLAC and APOLLON at ELFE.

\subsection{Collider}

Two colliders will govern the high energy part of spin physics in
future: the polarized proton collider RHIC at BNL~\cite{GotoSaito},
and possibly the HERA collider at DESY which may have polarized
protons in future~\cite{Hughes}.

In both machines, the acceleration and storage of polarized protons is
a major challenge to machine physicists. Several Siberian snakes will
be installed which compensate depolarizing resonances of the beam
polarization.

RHIC will start its physics program in 2000. Main points on the agenda
are the measurement of the antiquark polarization and the gluon
polarization. The antiquark polarization can be extracted using
Drell-Yan production via $W^+$ and $W^-$. As $W$-production depends on
flavor and on helicity, the experiment can extract $\Delta u$, $\Delta
\bar u$, $\Delta d$ and $\Delta \bar d$ separately.  The gluon
polarization is approached by the production of prompt photons,
$\pi^0$'s, jets and heavy quarks (charm).

The main aims of a polarized HERA collider are the measurement of the
spin structure functions at low $x$, which will improve the precision
of the verification of the fundamental Bjorken sum rule, and the
polarization of the gluon. As at RHIC, the gluon polarization can be
extracted from the production of heavy flavors and from jet
production.

\section{CONCLUSIONS}

Spin physics is a rich field. At this workshop an extraordinary number
of results was presented, from CERN, SLAC, Jefferson Lab, Mainz and
other institutions. A main source of results was the HERMES experiment
at DESY which released a variety of different analyses of data taken
in the last three years, some of which were exciting and unexpected.
Spin physics will continue to be a rich field when the future
facilities which were discussed at this workshop come into operation.

\section{ACKNOWLEDGMENTS}

I want to thank all speakers of the spin sessions for their excellent
contributions and I thank the organizers for the invitation and for
the perfect support of the session conveners.


\end{document}